 \documentstyle[12pt]{article}
 \newcommand{\be}{\begin{equation}}
 \newcommand{\ee}{\end{equation}}
 \newcommand{\ba}{\begin{eqnarray}}
 \newcommand{\ea}{\end{eqnarray}}

 \def\d{\partial}
\begin{document}
\title{The Birkhoff theorem for topologically massive gravity}
\author{Marco Cavagli\`a\thanks{E-mail: cavaglia@aei-potsdam.mpg.de; web
page: http://www.aei-potsdam.mpg.de/\~{}cavaglia}\\
\it
Max-Planck-Institut f\"ur Gravitationsphysik\\ \it
Albert-Einstein-Institut\\ \it
Am M{\"u}hlenberg 5\\ \it
D-14476 Golm, Germany}
\date{\today }
\maketitle

\begin{abstract}
We derive the general $\Sigma_2\times S$ solution of topologically massive
gravity in vacuum and in presence of a cosmological constant. The field
equations reduce to three-dimensional Einstein equations and the solution
has constant Ricci tensor. We briefly discuss the emergence of
non-Ricci flat solutions when spin is introduced. 

\medskip\noindent
Pac(s) numbers: 04.50.+h, 04.20.Cv, 04.20.Jb \\
Keyword(s): Alternative theories of gravity, Fundamental Problems,
Exact solutions  

\end{abstract}

\newpage

\noindent
Three-dimensional gravity allows for the construction of a
Chern-Simons-like term closely related to the four-dimensional
Pontryagin invariant \cite{DJT}. Adding this Chern-Simons term to the action
and varying the latter w.r.t.\ the metric, one obtains field equations
corresponding to Einstein equations plus a term proportional to a
symmetric, traceless, covariantly constant tensor of third derivative
order (Cotton tensor, $C_{\mu\nu}$) \cite{DJT}
\ba
&&G^\mu{}_\nu+{1\over m}C^\mu{}_\nu=\lambda\delta^\mu{}_\nu\,,
\label{tmg_eqs_updown}\\
&&C^\mu{}_\nu=\epsilon^{\mu\sigma\kappa}\nabla_\sigma
\left(R_{\nu\kappa}-{1\over 4}g_{\nu\kappa}R\right)\,,
\label{cotton_updown}
\ea
where $\epsilon$ is the completely antisymmetric Levi Civita tensor in
three-dimensions. The parameter $m$ in Eq.\ (\ref{tmg_eqs_updown}) has
mass dimension (in Planck units) and we have allowed for the existence of
a cosmological constant $\lambda$. This new theory is called topologically
massive gravity (TMG) and has several surprising properties \cite{DJT}. In
particular, it has no unitarity or ghost problems and appears to be
renormalizable. 

A quite large amount of attention has been devoted to the investigation of
the fundamental properties of TMG. In particular, TMG allows
for non-trivial (non Ricci flat) solutions in three dimensions \cite{sol}.
In this context, the study of exact solutions of TMG plays a major role. 
(Quoting Deser \cite{Deser}: ``[\dots] no-one has succeeded as yet in
finding the simplest possible, `Schwarzschild' solution to the nonlinear
model, {\it i.e.}, a circularly symmetric time-independent (we don't know
if there's a Birkhoff theorem) exterior geometry that obeys the [Eq.\
(\ref{tmg_eqs_updown})].'')

Having this in mind, the aim of this paper is the investigation of
cylindrically symmetric solutions of TMG. Given the

\begin{itemize}

\item{{\bf Definition 1:} A three-dimensional metric with topology
$\Sigma_2 \times S$
\be
ds^2=\gamma_{\mu\nu}(x)dx^\mu dx^\nu - \rho(x)^2 d\phi^2\,,
\label{metric_gen}
\ee
where $\phi\in[0,2\pi[$ and $\gamma_{\mu\nu}$ ($\mu,\nu=0,1$) is a generic
hyperbolic two-dimensional metric with signature $(1,-1)$, is {\it static}
when it possesses a Killing vector $\xi=\xi^\mu(x)\d_\mu$ (for the definition
of staticity in two-dimensional dilaton gravity see \cite{static});}

\item{{\bf Definition 2:} A solution of TMG equations is called {\it
locally Einstein trivial} if the Cotton tensor vanishes identically when
is evaluated on the solution;}

\end{itemize}
 
we prove the following 

\begin{itemize}

\item{{\bf Theorem:} The most general solution of Eq.\
(\ref{tmg_eqs_updown}) with topology $\Sigma_2 \times S$ is static and
locally Einstein trivial.}

\item{{\bf Corollary:} When spin is introduced, non-static and
non-Einstein trivial solutions do appear.}

\end{itemize}

The above theorem can be seen as the Birkhoff theorem for TMG. As a
consequence of the theorem, the most general solution of TMG with topology
$\Sigma_2 \times S$ is locally flat for $\lambda=0$ and locally De
Sitter/Anti-de Sitter for $\lambda\not=0$. As a result, TMG allows for 
the existence of BTZ black holes \cite{BTZ}. 

Conventions: Throughout the paper we use the signature $(+,-,-)$ and
Landau-Lifshits \cite{LL} conventions for the Ricci tensor. 

\begin{itemize}

\item{{\bf Proof:}

Let us consider the metric (\ref{metric_gen}). Since in two-dimensions any
metric is locally conformally flat we choose conformal light-cone
coordinates $u,v$ on $\Sigma_2$. Equation (\ref{metric_gen}) is cast in
the form
\be
ds^2=f(u,v)dudv - \rho(u,v)^2 d\phi^2\,,~~~~~~f(u,v)\ge 0\,.
\label{metric_conf}
\ee
Using (\ref{metric_conf}) the Einstein tensor $G^\mu{}_\nu$ and the Cotton
tensor read
\be
G^\mu{}_\nu=2\left(\matrix{{\cal H}_4&-{\cal H}_1/f&0\cr
-{\cal H}_2/f&{\cal H}_4&0\cr
0&0&{\cal H}_3}\right)\,,
\label{einstein_eval}
\ee
and
\be
C^\mu{}_\nu=\left(\matrix{0&0&2\rho{\displaystyle
2\d_u{\cal H}_1-f\d_v{\cal H}_3\over\displaystyle f^2}\cr\cr
0&0&-2\rho{\displaystyle 2\d_v{\cal H}_2-f\d_u{\cal H}_3\over\displaystyle
f^2}\cr\cr
{\displaystyle 2\d_v{\cal H}_2-f\d_u{\cal H}_3\over\displaystyle f\rho}
&-{\displaystyle 2\d_u{\cal H}_1-f\d_v{\cal H}_3\over\displaystyle f\rho}
&0}\right)\,,
\label{cotton_eval}
\ee
respectively, where
\ba
&\displaystyle
{\cal H}_1={f\over\rho}\d_v\left({\d_v\rho\over f}\right)\,,
~~~~~~~
&{\cal H}_2={f\over\rho}\d_u\left({\d_u\rho\over f}\right)\,,\\\\
&\displaystyle
{\cal H}_3={\d_u\d_v(\ln f)\over f}\,,
~~~~~~~
&{\cal H}_4={\d_u\d_v\rho\over f\rho}\,.
\ea
(Note that ${\rm det}\,(C^\mu{}_\nu)=0$.) From (\ref{tmg_eqs_updown}) and
(\ref{einstein_eval})-(\ref{cotton_eval}) it follows immediately
\ba
&&2\d_u{\cal H}_1-f\d_v{\cal H}_3=0\,,
\label{tmg-conditions1}\\
&&2\d_v{\cal H}_2-f\d_u{\cal H}_3=0\,.
\label{tmg-conditions2}
\ea
Thus the Cotton tensor vanishes and the space time is trivial. Equations
(\ref{tmg_eqs_updown}) can be written in the form
\ba
&&{\cal H}_1={\cal H}_2=0\,,
\label{tmg_eqs_eval1}\\
&&{\cal H}_3={\cal H}_4=\lambda/2\,.
\label{tmg_eqs_eval2}
\ea
The general solution of Eqs.\ (\ref{tmg_eqs_eval1})-(\ref{tmg_eqs_eval2}) is
\be
f(u,v)={d\rho\over dh}{dU\over du}{dV\over dv}\,,
~~~~~~
\rho(u,v)=\rho(h)\,,
\label{f-rho}
\ee
where $h(u,v)\equiv U(u)+V(v)$, $U(u)$, $V(v)$ are arbitrary functions.
Clearly, the solution possesses the Killing vector
\be
\xi= \left({dU\over du}\right)^{-1}\d_u-\left({dV\over dv}\right)^{-1}\d_v\,.
\ee
(Actually, in the case under consideration the space time is maximally
symmetric and possesses six Killing vectors.) The conformal factor $\rho$
satisfies the ordinary differential equation
\be
{d\rho\over dh}=C+{\lambda\over 4}\rho^2\,,
\label{conf_eval}
\ee
where $C$ is an integration constant. This concludes the proof of the
theorem.}

\end{itemize}

For sake of completeness, solving (\ref{conf_eval}) we have ($C\not=0$)
\ba
&&ds^2={C\over
\displaystyle\cos^2\left[{\sqrt{\lambda C}\over 2}(U+V)+K\right]}dUdV-
{4C\over\lambda}\tan^2\left[{\sqrt{\lambda C}\over 2}(U+V)+K\right]d\phi^2\,,
\label{Cnot0_1}\\
&&ds^2={C\over
\displaystyle\cosh^2\left[{\sqrt{-\lambda C}\over 2}(U+V)+K\right]}dUdV+
{4C\over\lambda}\tanh^2\left[{\sqrt{-\lambda C}\over 2}(U+V)+K\right]d\phi^2\,,
\label{Cnot0_2}
\ea
for $C\lambda>0$ and $C\lambda<0$ respectively. For $\lambda=0$ the
spacetime is locally flat, as we do expect from (\ref{f-rho}) and
(\ref{conf_eval}). Finally, when $C=0$ ($\lambda\not=0$) we obtain 
\be
ds^2={1\over\displaystyle\left[K-{\lambda\over 4}(U+V)\right]^2}
\left[{\lambda\over 4}dUdV-d\phi^2\right]\,,
\ee
that corresponds to De Sitter/anti-De Sitter space time in conformal
coordinates and is related to (\ref{Cnot0_1})-(\ref{Cnot0_2}) by a
coordinate transformation. 

The theorem can be (partially) extended to some matter-coupled models of
TMG. For instance, the Cotton tensor must necessarily vanish when a scalar
field is coupled to the system since the $(u,\phi)$ and $(v,\phi)$
components of the stress-energy tensor $ T^\mu{}_\nu$ are identically
zero. In this case, of course, Einstein triviality does imply neither
staticity nor Ricci-flatness. The solutions of TMG are the subset of the
solutions of Einstein equations that satisfy Eqs.\
(\ref{tmg-conditions1})-(\ref{tmg-conditions2}). 

As it has been suggested by Aliev and Nutku \cite{Nutku}, the theorem is
violated if spin is added to the model. We shall prove this by finding a
counterexample. The most general three-dimensional metric with a Killing
vector $\d_\phi$ can be locally cast in the form
\be
ds^2=F(u,v)^2 dudv + H(u,v)dud\phi + H(u,v)dvd\phi + E(u,v) d\phi^2\,.
\label{metric_gen2}
\ee
By a simple change of coordinates it is straightforward to see that the
ansatz (\ref{metric_gen2}) describes a stationary (spinning) geometry.
Indeed, setting $u=t+x$ and $v=t-x$ Eq.\ (\ref{metric_gen2}) is cast
in the form
\be
ds^2=F(x,t)^2\left[(dt+w(x,t)d\phi)^2-dx^2\right]-[F(x,t)^2w(x,t)^2-E(u,v)]
d\phi^2\,,
\ee
where $w(x,t)=H(x,t)/F(x,t)^2$. 

Let us consider for simplicity $E(u,v)=0$ and no cosmological constant.
Using Eq.\ (\ref{metric_gen2}) we find that the only non-vanishing
component of the Cotton tensor is $C^{\phi\phi}$. As a consequence, the
$(\mu,\nu)$ ($\mu,\nu=u,v$) components of Eq .\ (\ref{tmg_eqs_updown})
(with upper indices) coincide with three-dimensional vacuum Einstein
equations. They reduce to
\be
\left(\d_- H(u,v)\right)^2=0\,,
\label{eq_H}
\ee
where $\d_-=\d_u-\d_v$. The general solution of Eq.\ (\ref{eq_H}) is
\be
H(u,v)\equiv H(u+v)\,.
\label{H}
\ee
Substituting Eq.\ (\ref{H}) in the field equations the $(\mu,\phi)$
($\mu=u,v$) components of Eq.\ (\ref{tmg_eqs_updown}) are identically
satisfied. The $(\phi,\phi)$ component of the Cotton tensor is
\be
C^{\phi\phi}={H\over |HF|}\d_- G^{\phi\phi}\,,
\label{cotton-G}
\ee    
where 
\be
G^{\phi\phi}=-{1\over H^3F}\left(4H\d^2_{uv}F-\d_+ H\d_+F\right)\,,
\label{eq-G}
\ee
is the $(\phi\phi)$ component of the Einstein tensor and $\d_+=\d_u+\d_v$.

From Eq.\ (\ref{cotton-G}) it is straightforward to prove that if the
space time is static ($F\equiv F(u+v)$) it is also locally Einstein
trivial because $C^{\phi\phi}=0$.  However, non-static solutions do
appear. Let us choose $F\equiv F(u-v)$. Equation (\ref{eq-G}) becomes
\be
2F'''F-F''\left(2F'-m{H\over |HF|}F^3\right)=0\,,
\label{eq-F}
\ee
where $'$ means differentiation w.r.t.\ $u-v$. Let us consider for 
simplicity $H>0$. Integrating (\ref{eq-F}) we find
\be
2y''+myy'=Ay\,,~~~~y=|F|\,,
\label{eq-y}
\ee
where $A$ is an integration constant. When $A=0$ (\ref{eq-y}) can be
integrated explicitly. We obtain three different solutions
\ba
&&y_+={\alpha\over m}\tanh\left[{\alpha\over 4}(u-v)-\beta\right]\,,\\
&&y_-=-{\alpha\over m}\tan\left[{\alpha\over 4}(u-v)-\beta\right]\,,\\
&&y_0={\alpha/m\over\alpha(u-v)/4-\beta}\,.
\ea
The three-dimensional line elements are
\be
ds_i^2=y_i(u-v)^2 dudv+H(u+v)dud\phi+H(u+v)dvd\phi\,.
\label{metric-y}
\ee

Note that $ds_+^2$ and $ds_0^2$ are defined in the region
$\alpha(u-v)/4-\beta>0$ and $ds_-^2$ is defined in the regions
$\pi(2n-1)/2<\alpha(u-v)/4-\beta<n\pi$, where $n$ is a integer
number. (The regions of definition are reversed when $H(u,v)<0$.)

Clearly, Eqs.\ (\ref{metric-y}) describe a non-trivial non-static
geometry. Thus we have shown by an explicit example that non-static
solutions do occur in spinning cylindrically symmetric TMG geometries. The
presence of the Killing vector associated to cylindrical symmetry does
not forbid the existence of non-static solutions. 

\null\medskip

\noindent
{\large\bf Acknowledgments}

\noindent

This work was supported by a Human Capital and Mobility grant of the European
Union, contract  number ERBFMRX-CT96-0012. We are grateful to V.\ de Alfaro and 
Y.\ Nutku for useful discussions connected to the subject of this paper.


\thebibliography{999}

\bibitem{DJT}{S.\ Deser, R.\ Jackiw, and S.\ Templeton, {\it Phys.\ Rev.}
{\bf D 48}, 975 (1982); {\it Ann.\ Phys.} {\bf 140}, 372 (1982).}
 
\bibitem{sol}{I.\ Vuorio, {Phys.\ Lett.} {\bf B 163}, 91 (1985);\\
R.\ Percacci, P.\ Sodano, I.\ Vuorio, {\it Ann.\ Phys.} {\bf 176}, 344 (1987);\\
Y.\ Nutku, {\it Ann.\ Phys.} {\bf 195}, 16 (1989);\\
M.E.\ Ortiz, {\it Class.\ Quant.\ Grav.} {\bf 7}, 1835 (1990);
{\it Ann.\ Phys.} {\bf 200}, 345 (1990);\\
G.\ Cl{\'e}ment, {\it Class.\ Quant.\ Grav.} {\bf 7}, L93 (1990);
{\bf 9}, 2615 (1992); {\it ibid.} 2635 (1992)\\
B.\ Linet, {\it Gen.\ Rel.\ Grav.} {\bf 23}, 15 (1991);\\
Y.\ Nutku, {\it Class.\ Quant.\ Grav.} {\bf 10}, 2657 (1993).} 

\bibitem{Deser}{S.\ Deser, ``Dimensionally Challenged Gravities'',
Contribution to J.\ Stachel Festschrift, {\it preprint} gr-qc/9812013,
to appear in {\it Boston studies in the philosophy of science}.}

\bibitem{static}{A.T.\ Filippov, {\it Mod.\ Phys.\ Lett.} {\bf 11}, 1691
(1996); {\it Int.\ Jou.\ Mod.\ Phys.\ } {\bf A 12}, 13 (1997);\\
M.\ Cavagli\`a, {\it Phys.\ Rev.} {\bf D 57}, 5295 (1998);\\
T.\ Kl\"osch and T.\ Strobl, {\it Class.\ Quantum Grav.}
{\bf 13}, 965 (1996); {\bf 13}, 2395 (1997); {\bf 14}, 1689 (1997).}

\bibitem{BTZ}{M.\ Ba{\~n}ados, C.\ Teitelboim, and J.\ Zanelli {\it Phys. Rev.
Lett.} {\bf 69}, 1849 (1992).}

\bibitem{LL}{L.D.\ Landau and E.M.\ Lifshitz, ``The Classical Theory of
Fields'', Pergamon Press, 1962.}

\bibitem{Nutku}{A.N.\ Aliev, Y.\ Nutku, {\it Class.\ Quant.\ Grav.} {\bf
13}, L29 (1996).}
\end{document}